\documentstyle[multicol,osa,amstex]{revtex} 
\begin{document}

\setlength{\textwidth}{150mm}
\setlength{\textheight}{240mm}
\setlength{\parskip}{2mm}

\input{epsf.tex}
\epsfverbosetrue

\renewcommand{\baselinestretch}{1.0}

\title{Accurate switching intensities and length scales in 
quasi-phase-matched materials}

\author{Ole Bang, Torben Winther Graversen, and Joel F. Corney}

\address{Department of Informatics and Mathematical Modelling, 
Technical University of Denmark, DK-2800 Lyngby, Denmark}

\maketitle

\normalsize

\begin{abstract}
We consider unseeded Type I second-harmonic generation in quasi-phase-matched
(QPM) quadratic nonlinear materials and derive an accurate analytical 
expression for the evolution of the average intensity. 
The intensity-dependent nonlinear phase mismatch due to the QPM induced cubic 
nonlinearity is found. 
The equivalent formula for the intensity for maximum conversion, the crossing 
of which changes the nonlinear phase-shift of the fundamental over a period 
abruptly by $\pi$, corrects earlier estimates by more than a factor of 5.
We find the crystal lengths necessary to obtain an optimal flat phase 
versus intensity response on either side of this separatrix intensity.
\end{abstract}

\begin{multicols}{2}

\narrowtext
Since the observation of nonlinear phase-shifts in excess of $\pi$ through
cascading close to phase-matching \cite{BelGagIno89,SalHagSheSteStryVan92}, 
quadratic nonlinear or $\chi^{(2)}$ materials have been of significant 
interest in photonics \cite{SteHagTor96}.
With the maturing of the quasi-phase-matching (QPM) technique
\cite{ArmBloDucPer62}, in particular by electric-field poling of
ferro-electric materials, such as LiNbO$_3$ \cite{Fej98}, and by
quantum-well disordering in semiconductors \cite{Hel00}, the number of
possible applications of cascading in $\chi^{(2)}$ materials has
increased even more.

An effect of QPM gratings is to generate cubic nonlinearities in the 
equations for the average field due to non-phase-matched coupling 
between modes \cite{ClaBanKiv97}. 
This cubic nonlinearity appears in QPM with linear and/or nonlinear gratings 
\cite{CorBan00}, it can be focusing or defocusing, depending on the 
sign of the phase mismatch \cite{CorBan00}, and its strength can be 
significantly increased by modulating the grating \cite{BanClaChrTor99}.
Simulations of QPM systems have confirmed the presence of an intensity 
dependent nonlinear phase mismatch \cite{KobLedBanKiv98} and of soliton 
properties \cite{ClaBanKiv97,CorBan00}, that can only be described 
by including the cubic terms.
The Kerr nonlinearity is also known to distort the second-harmonic (SH) 
spectrum and introduce an intensity dependent nonlinear phase mismatch, 
as found both theoretically \cite{Akh72ChoBanCai91} and experimentally 
\cite{TelChi82}.
Thus, although the physical origin of the Kerr and the induced cubic 
nonlinearities is distinctly different, they have {\em qualitatively} 
similar effects on CW waves. 

In this letter we focus on the intensity dependent nonlinear phase-mismatch, 
which implies a finite separatrix intensity that can be used in efficient 
all-optical switching, as was shown for the Kerr nonlinearity in poled 
fibres \cite{ZhaTowSce95} and the QPM induced cubic nonlinearity 
\cite{KobLedBanKiv98}.
{\em A distinct feature of the averaged QPM model with induced cubic 
nonlinearities is that there is not a one-to-one correspondance between 
the physical field and the averaged field}.
The separatrix intensity obtained with no SH seeding in the averaged model
does therefore not accurately predict the real physical separatrix 
intensity of unseeded second-harmonic generation (SHG). 
We address this discrepancy and find the exact separatrix intensity for 
unseeded Type I SHG in QPM samples. 
We further show that the averaged model gives also quantitatively inaccurate 
results for no seeding of the average SH, since a basic assumption in the 
averaging procedure is violated, and we find the optimum crystal lengths for 
using the induced separatrix intensity for all-optical switching.

We consider a linearly polarized electric field $\vec{E}=\hat{e}
[E_1(z)\exp(ik_1z-i\omega t) + E_2(z)\exp(ik_2z-i2\omega t)+c.c.]/2$, 
propagating in a lossless QPM $\chi^{(2)}$ medium under conditions for 
type I SHG. The dynamical equations for the slowly varying envelopes take 
the form \cite{MenSchTor94Ban97}
\begin{eqnarray} 
   \label{phys1} 
   & & idE_1/dz + G(z)\chi_1E_1^*E_2 {\rm e}^{i\Delta kz}=0, \\
   \label{phys2} 
   & & idE_2/dz + G(z)\chi_2E_1^2{\rm e}^{-i\Delta kz}=0,
\end{eqnarray} 
where $E_1(z)$ is the fundamental wave (FW) with frequency $\omega$ and 
wavevector $k_1$, $E_2(z)$ is the SH with wavevector $k_2$, 
$\Delta k$=$k_2$$-$$2k_1$ is the wavevector mismatch, and 
$\chi_j$=$\omega d_{\rm eff}/(n_jc)$, with $n_j$=$n(j\omega)$ being the 
refractive index and $d_{\rm eff}$=$\chi^{(2)}/2$ being in MKS units.
The total intensity $I=\frac{1}{2}\eta_0(n_1|E_1|^2+n_2|E_2|^2)$ is conserved,
where $\eta_0$=$\sqrt{\epsilon_0/\mu_0}$ is the specific admittance of vacuum.
The $\chi^{(2)}$ susceptibility is modulated by the grating function 
$G(z)$ with unit amplitude and Fourier series 
$G(z)$=$\sigma\sum_ng_n{\rm e}^{in\kappa z}$, where $g_n$=0 for $n$ even and 
$g_n$=$2/(i\pi n)$ for $n$ odd.
This zero-average square-wave modulation is typical for QPM by domain 
inversion in ferro-electric materials, such as LiNbO$_3$. 
We define $\sigma$=sign$(\kappa)$ so that $G(z)$ is positive in the first 
domain. 

We further consider first order forward QPM with a short coherence 
length $L_d$$\sim$$L_c$$\ll$$L$, where $L$ is the crystal length, 
$L_d$=$\pi/|\kappa|$ is the domain length, and $L_c$=$\pi/|\Delta k|$ 
is the coherence length. 
Expanding the fields in Fourier series in the grating wavenumber $\kappa$, 
\begin{equation}
  \label{fourier}
  \hspace{-8mm}
  E_1=\sum_nw_n(z) {\rm e}^{in\kappa z} ,\;
  E_2=\sigma\sum_nv_n(z){\rm e}^{i(n\kappa-\beta)z},
\end{equation}
where {\em the harmonics are small compared to the dc-component}, a simple
first-order perturbation theory \cite{ClaBanKiv97,CorBan00} gives the 
dynamical equations for the slowly varying (on the scale of the domain 
length) average field 
\begin{eqnarray}
   \label{aver1}
   \hspace{-5mm}
   & & idw_0/dz + i\rho_1w_0^*v_0 + (\gamma_1|v_0|^2-\gamma_2|w_0|^2)w_0=0, \\
   \label{aver2}
   \hspace{-5mm}
   & & idv_0/dz + \beta v_0 - i\rho_2w_0^2 + 2\gamma_2|w_0|^2v_0=0,
\end{eqnarray}
where $\beta$=$\Delta k-\kappa$$\ll$$\kappa$ is the residual mismatch, 
$\rho_j$=$\chi_j2/\pi$, and $\gamma_j$=$\chi_j\chi_1(1-8/\pi^2)/\kappa$. 
Since $L_d$$\sim$$L_c$$\ll$$L$ the sign of $\kappa$ is the sign of the 
mismatch, sign$(\kappa)$=sign$(\Delta k)$. 
Thus the QPM induced cubic nonlinearity can be both focusing and defocusing,
depending on the sign of the mismatch $\Delta k$, just as the effective cubic 
nonlinearity obtained in the cascading limit \cite{SteHagTor96}. 
The harmonics are given by
\begin{equation}
  \label{harm}
  \hspace{-8mm}
  n\kappa w_{n\ne0} = \chi_1 g_{n-1} w_0^*v_0, \;\;\;
  n\kappa v_{n\ne0} = \chi_2 g_{n+1} w_0^2, 
\end{equation}

The average model (\ref{aver1}-\ref{aver2}) conserves the total average 
intensity $I_0=\frac{1}{2}\eta_0(n_1|w_0|^2+n_2|v_0|^2)$ and the quantity
$\epsilon=(1-u)\sqrt{u}\sin(\phi_2-2\phi_1)+Cu+Du^2$.
Here $\phi_1(z)$ and $\phi_2(z)$ are the phases of $w_0(z)$ and $v_0(z)$,
respectively, and $u(z)=\frac{1}{2}\eta_0n_2|v_0(z)|^2/I_0$ is the fraction
of average intensity $I_0$ in the average SH. 
The parameters are $D=\frac{3}{2}\gamma_1\sqrt{J}/\rho_1$ and 
$C=-\frac{1}{2}\beta/(\rho_1\sqrt{J})-\frac{4}{3}D$, where 
$I_0=\frac{1}{2}\eta_0n_2J$.
In the physical regime, where the polynomial
$f(u)=-D^2u^4+(1-2CD)u^3-(2+C^2-2\epsilon D)u^2+(1+2\epsilon C)u-\epsilon^2$
has four real roots $u_0\le u_1\le 1\le u_2\le u_3$, we solve 
Eqs.~(\ref{aver1}-\ref{aver2}) by quadrature and find the average 
SH intensity 
\begin{equation}
  \label{sol}
  \hspace{-8mm}
  u(z) = [\,u_3\,{\rm sn}^2(rz|m)+u_0p\,]/[\,{\rm sn}^2(rz|m)+p\,],
\end{equation}
where $p=(u_3-u_1)/(u_1-u_0)$, $mp=(u_3-u_2)/(u_2-u_0)$, and
$r=\frac{3}{2}|\gamma_1|J\sqrt{(u_2-u_0)(u_3-u_1)}$ are real 
positive parameters. 
The Jacobian Elliptic sn$(z|m)$ function is periodic with period 4K$(m)$, 
where K$(m)$ is the complete elliptic integral of the first kind 
\cite{AbrSte72}. 
The solution (\ref{sol}) is thus periodic with the period 2K$(m)/r$
and becomes aperiodic when $u_1$=$u_2$=1 ($m$=1), in which case 100\% 
transfer of power to the SH is predicted by the average model.
From $f(1)$=0 we obtain the separatrix $C+D$=$\epsilon$.

The solution (\ref{sol}) was found earlier for no seeding of the average SH, 
i.e.~with $u(0)$=$u_0$=$\epsilon$=0 \cite{KobLedBanKiv98}. 
In this case $I$=$I_0$ to lowest order and the separatrix $C$$+$$D$=0 
corresponds to the physical intensity $I$=$I_{\rm s}^0$=$\frac{1}{2}
\eta_0n_2J_{\rm s}^0$, where $J_{\rm s}^0$=$-\beta/\gamma_1$. 
It was further proven \cite{KobLedBanKiv98} that when the separatrix is
crossed the phase-shift of the FW over a period changed by exactly $\pi$, 
a result that is also valid for our general solution (\ref{sol}).
However, it was never investigated whether $I_{\rm s}^0$=$-\frac{1}{2}
\eta_0n_2\beta/\gamma_1$ was an accurate prediction of the separatrix 
intensity in the physical system (\ref{phys1}-\ref{phys2}). 

\begin{figure}
  \setlength{\epsfxsize}{8.0cm}
  \hspace{-3mm}\mbox{\epsffile{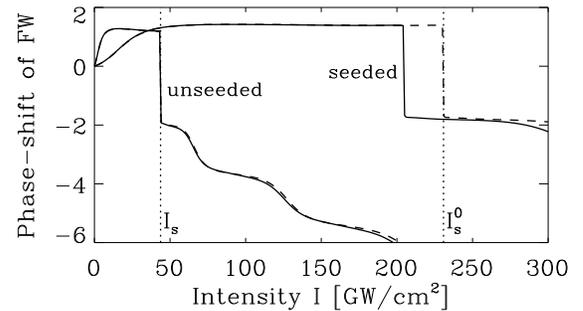}}
  \caption{Nonlinear phase-shift of the FW versus intensity $I$ for bulk 
  LiNbO$_3$, as predicted by the averaged model (\ref{aver1}-\ref{aver2}) 
  (dashed) and found numerically from the physical 
  Eqs.~(\ref{phys1}-\ref{phys2}) (solid). 
  The QPM domain length is $L_d$=8.8$\mu$m and the crystal length is
  $L$=351$\mu$m (174$\mu$m) for unseeded (seeded) SHG. The vertical 
  dotted lines mark the separatrices $I_{\rm s}$ and $I_{\rm s}^0$.}
  \label{phase}
\end{figure}

As a typical sample we consider bulk LiNbO$_3$, a FW wavelength of 
$1.064\mu$m, $d_{\rm eff}$=30pm/V, $n_1$=2.2, and $n_2$=2.23, which gives 
the coherence length $L_c$=8.9$\mu$m.
In Fig.~\ref{phase} we show the nonlinear phase-shift of the FW versus 
the physical intensity $I$ as predicted by the unseeded average model 
(\ref{aver1}-\ref{aver2}) with $u(0)$=$v_0(0)$=0 and found by numerical 
integration of the corresponding physical system (\ref{phys1}-\ref{phys2}) 
with initial condition determined from Eqs.~(\ref{fourier}) and (\ref{harm}). 
With the domain length $L_d$=8.8$\mu$m the predicted separatrix intensity
$I_{\rm s}^0$=231GW/cm$^2$ is 26GW/cm$^2$ too high compared to what is 
actually found in the physical system (\ref{phys1}-\ref{phys2}). 
This inaccuracy has not been observed before and is mainly due to the 
average SH being initially zero, which violates the requirement that 
the harmonics are small compared to the average field. 

At this point we stress that {\em no seeding in the average
model (\ref{aver1}-\ref{aver2}) implies a seeding of the physical SH}, as 
given by the transformations (\ref{fourier}) and (\ref{harm}), 
i.e.~$v_0(0)$=0 gives $E_1(0)$=$w_0(0)$ and $E_2(0)$=$-i(\sigma\rho_2/\kappa)
w_0^2(0)$. 
The experimentally relevant setup is unseeded SHG with $E_2(0)$=0, for which 
Fig.~\ref{phase} shows that the physical phase versus intensity curve is 
indistinguishable from the prediction of the average model ($v_0$ is now 
seeded and thus no assumptions are violated).
The physical separatrix intensity for unseeded SHG is 44GW/cm$^2$, i.e., 
5.3 times lower than the prediction $I_{\rm s}^0$ of the unseeded averaged 
model. Note that the averaged model without the cubic terms would not predict 
any separatrix.

From Eqs.~(\ref{fourier}) and (\ref{harm}) with $E_2(0)$=0 we find that 
$v_0(0)$=$i(\rho_2/\kappa)x^2$ where $x$=$w_0(0)$ is real and
$E_1(0)=x+(\rho_1\rho_2/\kappa^2)x^3$.
Then $\epsilon$=$n_2\rho_2\sqrt{J}/(n_1\kappa)$ to lowest order and thus 
$J_{\rm s}$=$-(1$$-$$8/\pi^2)\beta/\gamma_1$. 
This gives the physical separatrix intensity {\em for unseeded SHG}
\begin{equation}
  I_{\rm s} = -\frac{\eta_0n_2}{2}\left( 1-\frac{8}{\pi^2}\right)
              \frac{\beta}{\gamma_1}
\end{equation}
since again $I$=$I_0$ to lowest order. 
For $L_d$=8.8$\mu$m this gives $I_{\rm s}$=44GW/cm$^2$, which is exactly 
the numerically found value (see Fig.~\ref{phase}). 
Our numerical simulations confirm this accuracy for all values of the domain 
length satisfying $L_d\sim L_c\ll L$ and also for negative $\Delta k$.

A first requirement for the QPM-induced separatrix to be relevant for 
switching purposes is that the intensity $I_{\rm s}$ is low, 
which requires the ratio $\beta/\gamma_1$ to be small, i.e.~the effective 
mismatch must be small (but nonzero) and the induced cubic nonlinearity 
strong. 
We will not discuss more appropriate materials than LiNbO$_3$, but we note 
that the average model is very general \cite{BanClaChrTor99,CorBan00},
allowing $\gamma_1$ to be significantly increased by, e.g., modulating 
the QPM grating \cite{BanClaChrTor99} or by a strong dc-value of the 
grating \cite{CorBan00}.

\begin{figure}
  \setlength{\epsfxsize}{8.0cm}
  \hspace{-3mm}\mbox{\epsffile{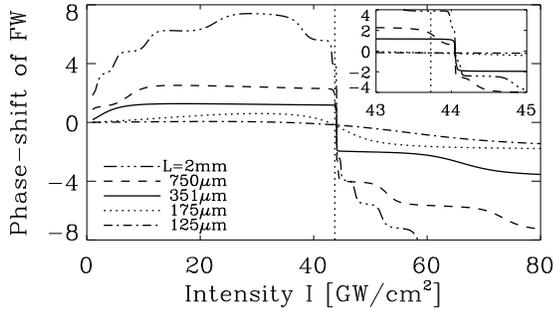}}
  \vspace{-4.5cm}
  \caption{Nonlinear phase-shift of the FW versus intensity $I$ for unseeded
  SHG in bulk LiNbO$_3$. The QPM domain length is $L_d$=8.8$\mu$m. The 
  crystal length is given in the figure.  The vertical dotted line marks 
  the separatrix $I_{\rm s}$=44GW/cm$^2$.}
  \label{length}
\end{figure}

So far the induced separatrix has not been observed in experiments. 
The reason could be:
(i) The intensity was too low; 
(ii) the induced and inherent self-phase modulation (SPM) terms eliminate 
each other - conventional $\chi^{(2)}$ materials are self-focusing and have 
normal dispersion and thus the inherent SPM coefficient is positive, while
the induced SPM coefficient ($-\gamma_2$) is negative;
(iii) the crystal length was inappropriate.
In Fig.~\ref{length} we illustrate the effect of the crystal length.
Clearly, the separatrix becomes more and more ``hidden'' in the overall 
variation due to the quadratic nonlinearity, and the flat plateaus on 
each side become narrower, as $L$ is increased from the optimal value 
around 351$\mu$m. When $L$ is too short the separatrix is lost.

\begin{figure}
   \setlength{\epsfxsize}{8.2cm}
   \hspace{-7mm}\mbox{\epsffile{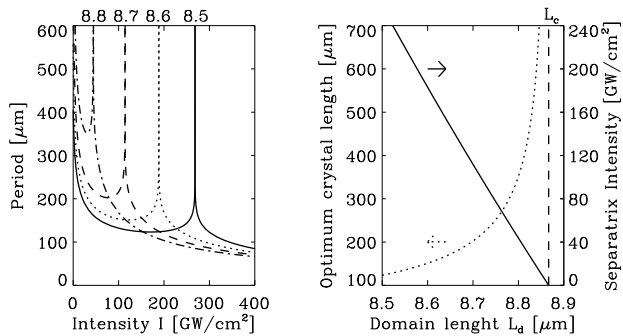}}
   \caption{Left: Period versus intensity $I$ for unseeded SHG in bulk
   QPM LiNbO$_3$ with the QPM domain length given in microns at the curves.
   Right: Optimal crystal length and separatrix intensity $I_{\rm s}$ versus 
   domain length.}
   \label{optimum}
\end{figure}

In Fig.~\ref{optimum} we show the period $2K(m)/r$ of the solution 
(\ref{sol}) versus intensity.  
The minimum period between $I$=0 and the separatrix $I$=$I_{\rm s}$ 
provides a good measure of the optimal crystal length observed in 
Fig.~\ref{length}, for which the $\pi$-shift due to the separatrix 
is the clearest and the flat phase versus intensity plateaus on either 
side are the broadest.
From Fig.~\ref{optimum} we see how this optimal length increases as the 
domain length approaches the coherence length 8.9$\mu$m, i.e.~as exact 
effective phase-matching is approached. 
At the same time the separatrix intensity decreases. 

Thus it is desirable to work close to, but not exactly at, exact
effective phase-matching in order to reduce the holding intensity 
in the switching process and keep a reasonable crystal length. 
Clearly the resolution in the photolithographic process could be 
an issue if the separatrix is to be measured and used.

The research is supported by the Danish Technical Research Council
through Talent Grant No.~5600-00-0355.

\end{multicols}

\end{document}